\documentclass{not_eas}
\usepackage{graphicx}
\usepackage{subfig}
\TitreGlobal{Virtual Spectro}
\runningtitle{a tool for Spectroscopic signatures of plasma}
\begin{document}

\title{Virtual Spectro, a tool for Spectroscopic signatures of plasma} 
\author{Michel Busquet}\address{ARTEP,inc, Ellicott City, MD - USA}
\begin{abstract}
We present here the tool "Virtual Spectro", a post-processor for spectroscopics analysis of hydrodynamic codes. We describe purpose and method of this tool, and the first application to analysis of high energy laser driven radiative shocks.
\end{abstract}
\maketitle
\section{Scaling down Radiative shocks from Astrophysical objects to laboratory experiments}
Radiative shocks (RS) are strong, supercritical shocks, where the temperature in the shock front is large enouch to emit photons that preheat the unshocked region and launch an ionizing radiative wave with a velocity larger than the shock velocity. RS are found in astrophysics in SN driven outwards shocks in the surrounding medium, in accretion shock and probably in jet driven bow shocks. In the laboratory, RS are launched either with blast wave (with point-like, instant energy deposition in tenuous gases) or with miniature shock tubes. In this case a high energy laser-ablated thin foil is accelerated in a cm-size tube filled with gas.
Existence of this radiative precursor (RP) requires a smaller shock velocity when the radiation mean free path is small, which comes with large atomic number. Therefore Argon or Xenon gas are preferred. A short bibliography of these experiments can be found in (Busquet \etal \cite{PALS1}). The RP can be understood as a Marshak wave, driven by radiation diffusion. However we found in a recent experiment that this paradigm is oversimplified, and that spectral features have to be included in the description.

\section{What we learned from previous experiment}
On the PALS laser facility, we recently performed a RS experiment which revealed the strange feature of a {\it split precursor} (Busquet \etal, \cite{PALS2}). Thanks to high optical quality targets (Busquet \etal, \cite{TARGET}), we were able to record RP front for more than 40 ns and to detect ionization plateau and subsequent secondary density jump build-up. We proposed that this split precursor is related to spectral opacities variations and spectral effects on radiative conduction. On the other hand, spectroscopics may be in future a powerful diagnostic of the RS. However the low resolution spectra obtained during this experiment were dominated by wall emission (Busquet, \cite{BAPS2011}), probably because temperature of the Xenon was not large enough. Detailed 2D post-processing radiative transfer is needed to sort out wall and bulk gas contributions. Also, wall X-ray re-emission, as measured on a dedicated Xe-jet experiment (Busquet \etal, \cite{ALBEDO}), has to be accounted for in such RS propagation studies. All these pushed us to update our spectroscopic post-processor and analyze hydrodynamic computation of RS in Xenon. 
\section{Virtual Spectro}
    To perform a spectroscopics analysis of the hydro-simulation, we use our post-processor code Virtual Spectro (Busquet \etal, \cite{VS2D}). VS can process results from 1D hydro-rad code such as EXMUL (Busquet, \cite{EXMUL}, Busquet \etal \cite{EXMUL2}), our extended version of the MULTI code (Ramis, \etal, \cite{MULTI}). It can also process results from 2D hydrocodes using snapshot dumps projected on quadrangular grids (Fig.1). It use precomputed spectral opacity {\it and} emissivity, either for all transitions in the LTE case (used here for this study), or for groups of  transitions in the non-LTE case (Busquet \etal, \cite{VS2D}). VS solves the radiative transfer for each directions of interest :
\[\frac{{\partial {I_\nu }(\overrightarrow \Omega  )}}{{\partial \overrightarrow r }}\,\,\, = \,\,{K_\nu } \times \left( {{S_\nu } - {I_\nu }(\overrightarrow \Omega  )} \right)\]
where  {\overrightarrow r } is the unit vector along the direction  {\overrightarrow \Omega} using the integral solution 
\[
\begin{array}{l}
 I_\nu  (x) = I_\nu  (x_0 )\,{\rm{e}}^{{\rm{ - }}\tau _\nu  {\rm{(x}}_{\rm{0}} {\rm{,x)}}}  + \int_{x_0 }^x {S_\nu  (t)\,{\rm{e}}^{{\rm{ - }}\tau _\nu  {\rm{(t,x)}}} \,\kappa _\nu  (t)\,\,} dt \\ 
 \tau _\nu  {\rm{(y,x) = }}\int_y^x {\,\kappa _\nu  (t)\,\,} dt \\ 
 \end{array}
\]
  In VS, coupling (i.e. retro-action) of radiation with time dependent atomic population is restricted to ions with less than 6 bound electrons, like C in (Weaver \etal, \cite{VS1D}).

\begin{figure}
 \centering
\includegraphics[width=7.2cm] {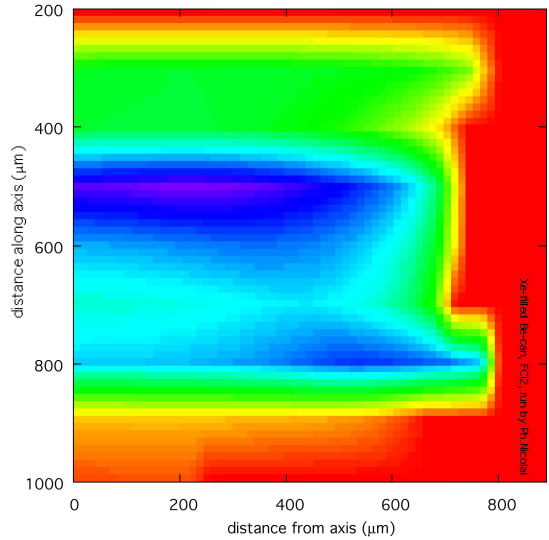}
\caption{2D map of Xe emission in laser heated holhraum (Busquet \etal, \cite{VS2D})}
\label{Be-can}
\end{figure}
\begin{figure}
 \centering
$\begin{array}{c@{\hspace{0.0cm}}c}
\multicolumn{1}{l}{\mbox{\bf  }} &
	\multicolumn{1}{l}{\mbox{\bf  }} \\ [+0.0cm] 
  \includegraphics[width=6 cm] {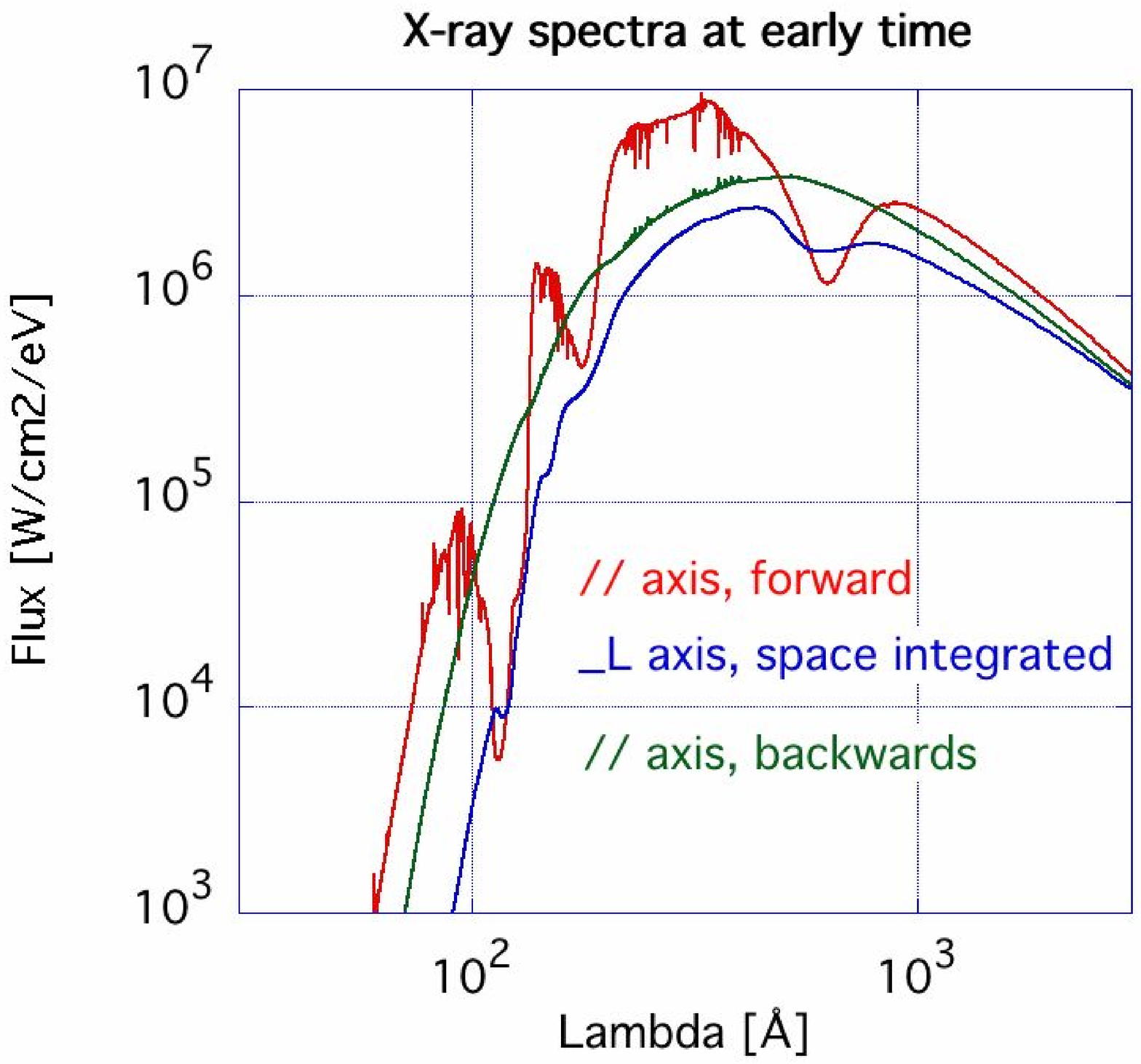}
  \includegraphics[width=0.5cm] {white.eps}
  \includegraphics[width=6 cm] {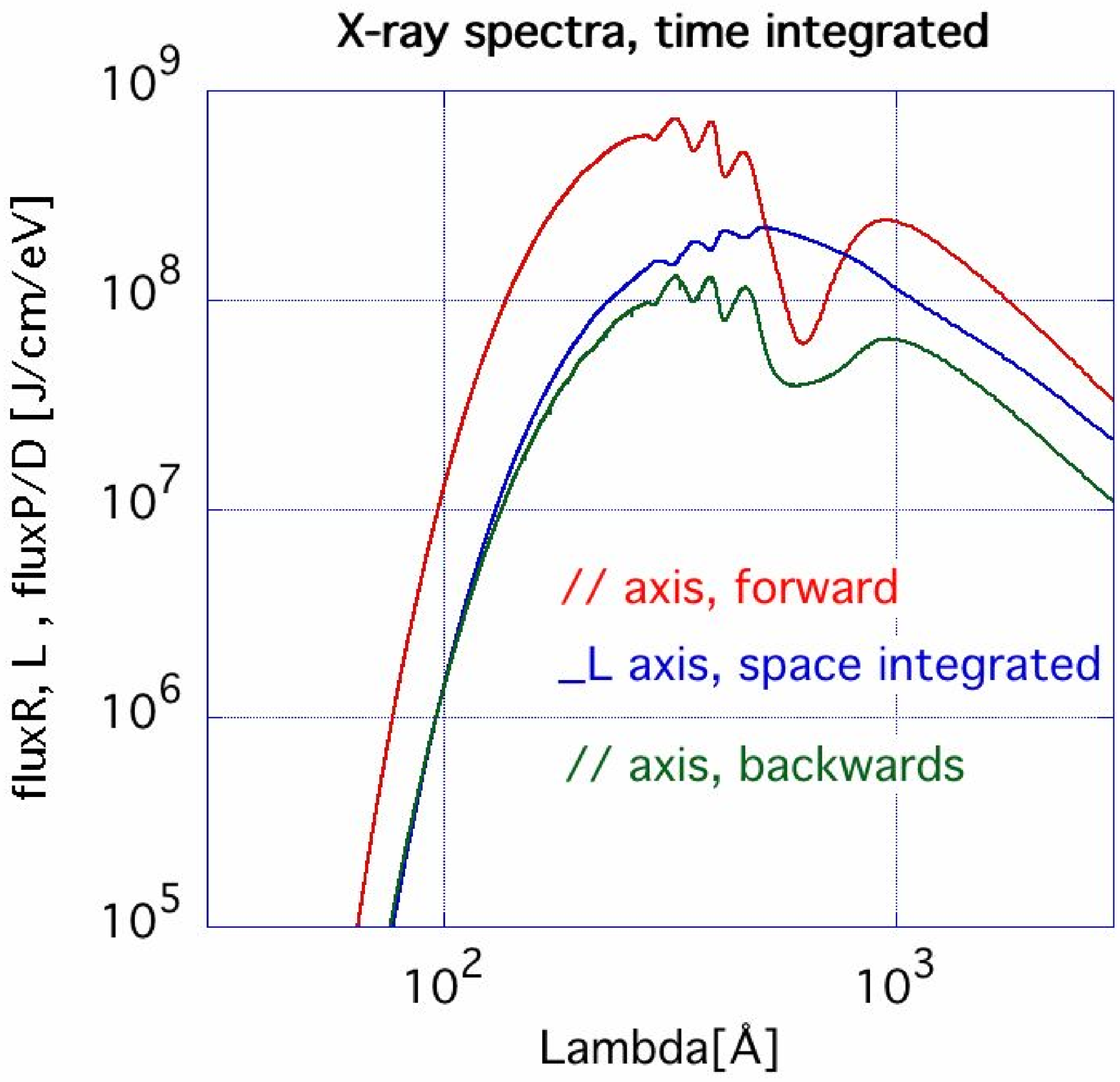}
\end{array}$
\caption{early time(left) and time integrated(right) synthetic spectra}
\label{spectra}
\end{figure}
To be used in VS, the opacities and emissivities can be computed with the non-LTE superconfiguration code SCROLL (Bar-Shalom \etal, \cite{SCROLL}) as in Fig.1, with the detailed level atomic structure code HULLAC-v9 (Klapisch \etal, \cite{HULLAC2}, Busquet \etal, \cite{HULLAC}),  or with the recently revised version of the LTE superconfiguration code STA. For these RS studies, we use LTE opacities from STA in the density range 0.1-100 mg/cm3 and the temperature range 1-49 eV. Note that only a few high quality opacity code is able to go down to such low temperature. Typical PALS laser condition (around $10^{14} W/cm^2$ in blue,  at  $\lambda$  =0.438$\mu$m, 150 J in 0.35 ns) has been used as parameters for a EXMUL simulation, with LTE opacities from STA, accounting for lateral radiative losses with an albedo of 0.4. Using VS, user can choose 
- the direction of observation, 
- the time, frequency and space ranges, 
- the time, frequency and space resolution or integration
- the transfer function of the recording system and filters
Some of the possibilities of VS for RS studies are illustrated in Fig.2. One can observe from these spectra the influence of time integration, and of direction of observation.

Comparison of outgoing spectra with various hypothesis of the hydro-rad simulations (like grey or multigroup approximations) are in progress. Variation with space will  be computed once the 2D hydro simulations are completed.
\section{Spectroscopic signatures}
VS is able to compute synthetic time, space and spectral, resolved or integrated, recording and simulate any possible spectrocopic diagnostic such as space resolved emission or absorption spectral measure, 2D map of monochromatic absorption, time $\times$ wavelength or time $\times$ space images, ... It will be a valuable tool for experiments analysis.

Ê

\footnotesize{
Support to this work comes from : NRL-laser plasma branch, LASERLAB-Europe, Observatoire de Paris, ARTEP,inc, R.S.I, CEA, and CRASH center.
}

\end{document}